\documentclass[journal]{vgtc}                

\ifpdf
  \pdfoutput=1\relax                   
  \usepackage{graphicx}                
  \DeclareGraphicsExtensions{.pdf,.png,.jpg,.jpeg} 
\else
  \usepackage{graphicx}                
  \DeclareGraphicsExtensions{.eps}     
\fi%
\graphicspath{{figures/}{pictures/}{images/}{./}} 
\usepackage{multirow}
\usepackage{microtype}                 
\PassOptionsToPackage{warn}{textcomp}  
\usepackage{textcomp}                  
\usepackage{mathptmx}                  
\usepackage{times}                     
\usepackage{cite}                      
\usepackage{tabu}                      
\usepackage{booktabs}                  
\usepackage[normalem]{ulem}
\usepackage{enumitem}
\usepackage{amsmath,amsfonts,amssymb,amsthm, bm}
\usepackage{fontawesome}

\usepackage[switch]{lineno}
\usepackage[section]{placeins}
\usepackage{svg}

\usepackage{xcolor}
\definecolor{ao-blue}{HTML}{2c7fb8}

\definecolor{sm-green}{HTML}{228c22}

\definecolor{sm-red}{HTML}{FF0000}
\newcommand{\red}[1]{{\color{sm-red}{#1}}}
\newcommand{\green}[1]{{\color{sm-green}{#1}}}

\DeclareMathAlphabet\mathbfcal{OMS}{cmsy}{b}{n}

\onlineid{1198}

\vgtccategory{Research}
\vgtcpapertype{algorithm/technique}

\title{
Competing Models: Inferring Exploration Patterns and Information Relevance via Bayesian Model Selection
}

\author{Shayan Monadjemi, 
Roman Garnett, and 
Alvitta Ottley}
\authorfooter{
\item
 Shayan Monadjemi is with Washington University. monadjemi@wustl.edu.
\item
Roman Garnett is with Washington University. garnett@wustl.edu.
\item
 Alvitta Ottley is with Washington University. alvitta@wustl.edu.
}

\shortauthortitle{Monadjemi \MakeLowercase{\textit{et al.}}: Competing Models: Inferring Exploration Patterns and Information Relevance via Bayesian Model Selection}

\abstract{
   Analyzing interaction data provides an opportunity to learn about users, uncover their underlying goals, and create intelligent visualization systems. The first step for intelligent response in visualizations is to enable computers to infer user goals and strategies through observing their interactions with a system. Researchers have proposed multiple techniques to model users, however, their frameworks often depend on the visualization design, interaction space, and dataset. Due to these dependencies, many techniques do not provide a general algorithmic solution to user exploration modeling. In this paper, we construct a series of models based on the dataset and pose user exploration modeling as a Bayesian model selection problem where we maintain a belief over numerous competing models that could explain user interactions. Each of these competing models represent an exploration strategy the user could adopt during a session. The goal of our technique is to make high-level and in-depth inferences about the user by observing their low-level interactions. Although our proposed idea is applicable to various probabilistic model spaces, we demonstrate a specific instance of encoding exploration patterns as competing models to infer information relevance. We validate our technique's ability to \textit{infer exploration bias}, \textit{predict future interactions}, and \textit{summarize an analytic session} using user study datasets. Our results indicate that depending on the application, our method outperforms established baselines for bias detection and future interaction prediction. Finally, we discuss future research directions based on our proposed modeling paradigm and suggest how practitioners can use this method to build intelligent visualization systems that understand users' goals and adapt to improve the exploration process.
} 

\keywords{User Interaction Modeling, Bayesian Machine Learning}

\CCScatlist{ 

}

\vgtcinsertpkg

\begin{document}


\firstsection{Introduction}

\maketitle

Visual analytics has transformed the process of reasoning with data by viewing humans and machines as teammates with unique strengths. Due to the growing complexity and volume of data, there is a need for more intelligent visual analytic systems to assist in data exploration, decision-making, and communication of findings. 
Studies suggest that even with interactive visualizations, there are human-related factors that hinder data exploration and informed decision-making. 
For instance, analysts often search through a large amount of information where the irrelevant portions of the data can be distracting and exceed the limits of human cognition \cite{halford2005many}. An intelligent visual analytic system can respond by only showing portions of the data relevant to the analyst. An everyday user might interact with only a biased subset of the data which impedes decision-making \cite{dimara2016attraction}. An intelligent visual analytic system can respond by informing the user about their biases through notifications or mitigate their biases through modifying the visualization. Creating an effective intelligent visual analytic system requires careful design considerations including the level of intrusiveness and the means of intervention\cite{wall2019toward}. Regardless of what form this \textit{intelligent visual analytic system} takes on, it must be able to offer well-informed assistance to the user.

One promising approach to informing useful machine response is through capturing and analyzing users' interaction data. Research in the area of \textit{Analytic Provenance} has shown that interaction data can reveal valuable information about the user and their analysis strategies and how the machine teammates can better assist the user.

For example, Dabek and Caban proposed a grammar-based approach to uncover common patterns among a group of users \cite{dabek2016grammar}, resulting in suggestions to assist users in data exploration.  Battle et al. \cite{battle2016dynamic} used a Markov chain that utilized navigation behavior to determine portions of a satellite image that a given user will likely explore in the future, resulting in improving latency by 430\%. Similarly, Ottley et al. \cite{ottley2019follow} and Wall et al. \cite{wall2017warning} used hidden Markov models to infer users' attention and cognitive bias respectively, resulting in anticipating future clicks and analyzing user exploration bias. For a more comprehensive survey on user modeling with provenance data, see Ragan et al.\cite{ragan2015characterizing} and  Xu et al. \cite{xu2020survey}. Unfortunately, much of the existing solutions depend on the visualization encoding or the interaction design, and do not easily provide a general algorithmic solution to uncovering exploration patterns and quantifying information relevance.

In this paper, we take a more comprehensive view of user modeling that encompasses  what the user has done, is doing, and will do in the future. Specifically, we use observed interactions as model evidence, and we leverage Bayesian model selection to reason about the subset of a multi-dimensional dataset and the collection of dimensions that most likely gave rise to the observations.
This approach provides a valuable framework to build, maintain, and select models that we derive from the dimensions in the dataset, and it offers two key advantages. First, it utilizes straightforward probability theory that is easy to understand and implement. Second, a unique benefit of this approach is that we can make high-level inferences about a user's exploration pattern by passively observing low-level interactions with the data points. Once we identify the data dimensions that are most relevant, we demonstrate that we can use our models to achieve a variety of user modeling objectives: \textit{infer exploration bias}, \textit{predict future interactions}, and \textit{summarize an analytic session}. 
We validate our proposed framework using three crowdsourced test datasets and we compare our algorithm's performance to existing baseline models. 
We show that our modeling technique outperforms established baselines in most cases. Furthermore, we discuss the possibilities this modeling paradigm opens for future investigations.
We summarize our contributions as follows:
\begin{itemize}[nosep]
    \item \textit{We introduce a straightforward approach to inferring multi-dimensional data exploration strategy in real-time.}  We encode different exploration strategies as a set of models and utilizes Bayesian model selection to reason about which strategy is likely to give rise to past interactions with a point-based visualization.
    \item \textit{We demonstrate our method's flexibility to address multiple aspects of user explorations modeling when applied to information relevance.} Specifically, we show that we can apply this framework to infer exploration bias, to predict future interaction, and to summarize an analysis session.
    \item \textit{We validate the proposed framework using three crowdsourced user interaction datasets.} Our results indicate that, depending on the application, our method outperforms established baselines.
\end{itemize}

\newpage
\section{Background} \label{background}

Learning and modeling user behavior is a common goal in the Visual Analytics community \cite{ribarsky2016human, ragan2015characterizing, wall2017warning, ottley2019follow}. \emph{Analytic Provenance}, which broadly refers to the research goal of tracking and modeling the analysis process, provides valuable information for inferring certain characteristics of exploration. Ragan et al. \cite{ragan2015characterizing} categorize provenance into five categories: provenance of data, visualization, interactions, insights, and rationale. Furthermore, they categorize the primary purpose of tracking provenance data as \textit{recall}\cite{  dou2009recovering, lipford2010helping,simmhan2006performance}, \textit{replication}\cite{ del2007probe,  kadivar2009capturing, maguire2012taxonomy, north2011analytic}, \textit{action recovery}\cite{ derthick2001enhancing, dobovs20143d,groth2006provenance,kurlander1988editable,shrinivasan2008supporting},
\textit{collaborative communication}\cite{callahan2006vistrails, dunne2012graphtrail, ellkvist2008using, mahyar2014supporting}, \textit{presentation}\cite{heer2012interactive, heer2008graphical}, and \textit{meta-analysis}\cite{parker1995scirun, hensley2014provenance}.
This paper aims to leverage \textit{interaction provenance} data to learn about the users and their goals in order to improve the visualization system.

\subsection{User Modeling in Visualization}

Recent work has begun to investigate how learning about the users' analytic process can help us improve visualization systems.
For example, Battle et al. \cite{battle2016dynamic} demonstrated this concept by using user exploration modeling to intelligently pre-fetch data for a map visualization. They observed user actions such as \{\emph{hover, click, pan-left, pan-right, ...}\}, maintained a Markov chain model, and ranked the data for pre-fetching according to the likelihood of the users taking actions corresponding to said data. Their approach improved the overall latency by 430\%. 

Brown et al. \cite{brown2012dis} have proposed \emph{Dis-function} as a method to learn expert knowledge through interaction data as a distance function representing the underlying similarities between data points. In their framework, the expert user interacts with the visualization by positioning similar points close to one another. By observing the user's organization of points, the algorithm infers a distance function that mimics the experts' knowledge. In similar work, Iwata et al. \cite{iwata2013active} took a step further to minimize the number of user interactions needed for finding the desired visualization representing high dimensional data. Using an active learning approach, they queried the user to relocate points strategically in order to reduce the uncertainty of the visualization.

Dabek and Caban \cite{dabek2016grammar} have proposed a grammar-based approach to model the sequence of low-level interaction in a session. Their work addresses the overwhelming number of controls and interaction channels available on modern visual analytics systems by using finite state machines to make suggestions for future interactions during a session. While their technique is task-dependent and relies on training data from past users performing the same tasks, their findings show that appropriate recommendations help users in data exploration.

This set of investigations provide examples on how user exploration modeling can improve user experience and decision making outcomes in visual analytic systems. However, many of them suffer from dependence on visualization design, tasks, or datasets. In line with our motivation of building intelligent visual analytic system who observe user interactions and infer what the user has done, is doing, and will do in the future, we propose an algorithmic and generalizable approach to building competing models on how users may interact with data. Moreover, we use Bayesian iterative updating to maintain our beliefs during sessions. We specifically design our models to take a comprehensive view on interactive sessions by inferring exploration bias, predicting future interactions, and summarizing sessions. In the remainder of this section, we discuss the most closely related work to this paper.

\subsection{Inferring Exploration Bias}

In the context of visual analytics, \textit{bias} may arise from different sources. Related literature has studied the phenomena of bias from the lens of \textit{data, models, and users}. Gotz et al. \cite{gotz2009characterizing} proposed a toolkit to combat selection bias, where the source of bias is in the \textit{data}. For instance, filtering a real estate data set to include houses less than \$300K will automatically introduce a bias towards the location of selected houses since price and neighborhoods are highly correlated. Cabrera et al. \cite{cabrera2019fairvis} proposed a visual analytic tool, FairVis, to combat bias in machine learning \textit{models}. For example, a model designed for recommending employment candidates for interviews may unfairly disregard candidates solely due to their historic under-representation in a career field.

In this work, we are interested in bias from the lens of \textit{users}. As the most relevant work to this paper, Wall et al. \cite{wall2017warning} introduced metrics to measure cognitive bias through user interaction with visualized data. Cognitive bias is measured in terms of how much of the dataset has been explored, which attributes of the data points were important to the user, and what value of those attributes have influenced the users' decisions. As an example, imagine a parent who is interacting with a map of crimes to determine if a specific neighborhood is safe for their family. They may begin by looking for sex crimes in that specific neighborhood. The algorithm soon determines that the \emph{location} and \emph{type} of the crime are the two attributes driving the user's exploration, while the \emph{time} attribute has been uniformly explored. 
Presenting this bias to the user during the session may result in them confirming their interest in a subset of data, or it may encourage more exploration and mitigate unintentional bias. While mitigating information bias requires careful design considerations which are beyond the scope of this paper (e.g. degree of guidance, level of intrusiveness, and means of interference \cite{wall2019toward}), we demonstrate our technique's ability to infer intersectional bias in exploration which provides critical information for bias mitigation.

\subsection{Predicting Future Interactions}

The term \textit{interaction} can take on numerous meanings in the context of interactive visualizations. For example, \textit{interaction} may refer to low level events such as clicks, hovers, and drag/drops, or it may refer to higher level tasks such as filtering and sorting. As a result, Gotz et al. \cite{gotz2009characterizing} have proposed a taxonomy where analytic behavior is categorized in to four groups: \textit{tasks, sub-tasks, actions, and events}. Their taxonomy created a spectrum based on semantics, where \textit{tasks} on one end of the spectrum refer to interactions with high semantics (e.g. identify market insights for promising investments) and \textit{events} on the other end of the spectrum refer to interactions with poor semantics (e.g. an individual click).

In the current literature, \textit{predicting future interactions} can take on different meanings as well. For example, Dabek and Caban \cite{dabek2016grammar} build models to predict future \textit{low-level events}, whereas Ottley et al. \cite{ottley2019follow} and Battle et al. \cite{battle2016dynamic} build models to predict which \textit{data point} the user is likely to interact with next. Our work is related to the latter category, as we will infer the relavance of data points to the user in light of past visited data points and compare our technique with Ottley et al. \cite{ottley2019follow} as a baseline. Ottley et al. \cite{ottley2019follow} proposed a hidden Markov model (HMM) approach to maintain a belief over users' evolving attention and actions in a visualization system. They encode user clicks as a sequence of visual attributes with an added bias metric to determine what attribute is influencing the exploration. As user clicks arrive, they update the model and use particle filtering to infer a set of top-\emph{k} point candidates for the next click. In contrast to their work, our models are built on the \emph{data} rather than the \emph{visual elements}. Since not all attributes of a multi-dimensional dataset can have a visual representation, we believe that building the model on the dataset itself opens up an opportunity for learning about users' interest in \emph{any} of the attributes, not only the ones visualized.

\subsection{Summarizing Analytic Sessions}

The body of work in this area ranges from primitive undo/redo functionalities to communicating provenance data from collaborative sessions. Heer et al. \cite{heer2008graphical} explore tools for visualizing provenance data, and discuss how such tools can be used to improve user interface design. Xu et al. \cite{xu2018chart} propose \textit{Charts Constellations}, a visualization tool to aggregate provenance data from multiple analysts in order to communicate the depth of exploration in different parts of a dataset. Chen et al. \cite{chen2010click2annotate} observe low-level interactions and create annotations to assist users with insight externalization. Gratzl et al. \cite{gratzl2016visual} propose \textit{CLUE}, a modeling framework that joins exploration and communication of discoveries by automatically storing provenance data during exploration and generating a session report. 

In another related work, Sarvghad et al. \cite{sarvghad2016visualizing} designed an experiment where users explore a multi-dimensional dataset to gain insight into business performance of an online retailer. Their study suggests that those who were provided with a summary on which attributes they have covered in their exploration were able to formulate more questions and explore more broadly without sacrificing the depth of insights.
Our proposed model in this paper allows us to accomplish a similar premise as prior work: to summarize a session. This capability can simply communicate our model's understanding of the user or it can be presented to the user to promote more exploration in a similar fashion as Sarvghad et al. \cite{sarvghad2016visualizing}.

\section{Preliminary Definitions}

We begin with some definitions to frame the reader's understanding of the problem space. We assume the user interacts with a set of point-based visual data and each point of interest $\vec{x}$ can be described by a vector of $d$ continuous or discrete attributes. We define a corresponding $d$-dimensional \textbf{data space} of possible realizations of these attributes, $\mathcal{X} = \{X_1 \times X_2 \times ... \times X_d\}$, where $X_i$ is the  domain of the $i$th attribute. We will consider a visualization of a \textbf{dataset} -- that is, a collection of points in the data space -- $\mathcal{D} = \{\vec{x}_1, \vec{x}_2, ..., \vec{x}_m\}$, $\vec{x}_i \in \mathcal{X}$. We assume that a user generates a stream of \textbf{interaction data}, a sequence of items in the dataset the user has interacted with, which we will notate with $\mathcal{C} = \{\vec{c}_1, \vec{c}_2, ... \}$, $\vec{c}_j \in \mathcal{D}$. While our definition of interaction data is independent of the type of interaction, this paper considers examples where the interactions occur by \textit{clicking} or \textit{hovering}. Our technique, however, may be expanded to include datapoints from other means of interaction subject to limitations discussed in Section \ref{discussion}.
We approach user interaction modeling via the framework of Bayesian model selection. We construct competing high-level models of a user's exploration pattern based on different attributes of $\mathcal{D}$, and use a given set of interaction data to determine which of these models is the most plausible. Here, a model is a parametric family of probability distributions to explain a given set of observations, and we will construct models that allow for a variety of different subspaces/subsets of the data space to be relevant (or not) for a user, spanning a large space of hypotheses. In the Bayesian approach, there are three major steps to probabilistic inference:
\begin{itemize}[nosep]
    \item The \textbf{prior} represents a belief distribution over possible values of the parameter before data is observed. 
    \item The \textbf{likelihood} specifies the probability of observing a given set of observations (such as a click stream) assuming given values for the parameters.
    \item The \textbf{posterior} is the updated distribution over possible values of the parameter in light of the prior beliefs and the information in the observations.
\end{itemize}

Going forward, we initiate the \textit{prior} belief and maintain the \textit{posterior} belief over a set of competing models in Section \ref{subsec:maintaining-a-belief-over-models}. Moreover, we discuss the \textit{likelihood}, or the relevance of data points in light of past observations, in section \ref{subsec_info_rel}. Finally, we use the competing models to detect exploration bias, predict future interactions, and summarize a session (Section \ref{sec:high_level_insight}).

\begin{table}[!t]
\caption{Table of notations used in this paper and their descriptions.}
\begin{tabular*}{\linewidth}{ll}
\toprule
Notation & Description \\ \midrule
        $\mathcal{D}$          &      The set of all data points visualized                \\
        $\mathcal{C}$          &      The set of data points interacted with                \\
        $\mathcal{X}$          &      The space of all possible data points                \\
        $\mathbb{M}$           &      The set of all models          \\
        $\mathcal{M}_i$         &      An individual model from $\mathbb{M}$          \\
        $\mathbb{A}_{c, i}$         &      The set of continuous attributes relevant to $\mathcal{M}_i$          \\
        $\mathbb{A}_{d, i}$         &      The set of discrete attributes relevant to $\mathcal{M}_i$         \\
        \bottomrule
\end{tabular*}
\vspace{1mm}
\label{notation_table}
\vspace{-5mm}
\end{table}

\section{Competing Models Framework}

\label{sec_competing_model_framework}

The novelty of this work is that we present different exploration patterns as models, and use the Bayesian model selection framework to maintain a belief over all possible exploration patterns. 
Our approach to user interaction modeling assumes that users interact with visualizations with some (possibly subconscious) pattern in mind. The ultimate goal of our technique is to detect and model these exploration patterns. We define \textbf{model space} to be a finite set of models representing various exploration patterns.
The model space is denoted by \mbox{$\mathbb{M} = \{\mathcal{M}_1, \mathcal{M}_2, ..., \mathcal{M}_n\}$} where $\mathcal{M}_i$ is an individual model representing one possible exploration pattern.
Depending on the choice model space, these individual models could inform us about user's bias in certain attributes of data, a specific task a user may be performing, or something about a user's personality. In this work, we encode exploration patterns in terms of which subset of data dimensions drive user exploration. For a dataset with $d$ dimensions, we create a model space of $2^d$ models, where each model represents exploration based on a subset of $d$ dimensions. 
As a user interacts with a visualization, we maintain a belief over viability of each model given the observed interactions. The belief over our models is incrementally updated using the Bayes' rule:
\begin{equation}
    p(\mathcal{M}_i \mid \text{interactions}) \propto p(\text{interactions} \mid \mathcal{M}_i) p(\mathcal{M}_i),
\end{equation}
where each model is evaluated in light of observed interactions and prior beliefs. Next, we discuss the details of initiating a prior belief over the model space and updating the belief as the user interacts with data points.

\subsection{Maintaining a Belief over Competing Models} \label{subsec:maintaining-a-belief-over-models}

Before observing any interactions, our \textbf{prior} assumes a uniform distribution over the set of all possible models. That is, for every model, $\mathcal{M}_i$,
\begin{equation}
    p(\mathcal{M}_i) = \frac{1}{2^d}
\end{equation}
This choice of prior reflects the idea that we are uniformly uncertain at the beginning about what combination of attributes best describe exploration patterns. In other words, we consider every exploration pattern to be equally likely before observing any interactions. In Section \ref{boardrooms-analysis}, we will demonstrate an alternative prior that penalizes models which encompass exploration bias towards a large number of attributes.

As new interactions occur, we update these models throughout a session via the process of iterative Bayesian updating. That is, given the set of interactions, $\mathcal{C}$, the \textbf{posterior} belief over the set of models is:
\begin{equation}
    p(\mathcal{M}_i \mid \mathcal{C}) \propto p(\mathcal{C} \mid \mathcal{M}_i) p(\mathcal{M}_i)
    \label{eq_model_posterior}
\end{equation}
\noindent where $p(\mathcal{C} \mid \mathcal{M}_i)$ is the \textbf{likelihood} representing how well a model $\mathcal{M}_i$ can explain the observations in $\mathcal{C}$. We can expand the likelihood function via chain rule to get:
\begin{align}
    \begin{split}
        p(\mathcal{C} \mid \mathcal{M}_i) &= \prod_{\vec{c}_j \in \mathcal{C}}p(\vec{c}_j \mid \mathcal{M}_i, \mathcal{C}_{1:j-1}) 
    \end{split}
    \label{eq_model_likelihood}
\end{align}
In Eq.\ \ref{eq_model_likelihood}, $\vec{c}_j$ is the $j$th click and $\mathcal{C}_{1:j-1}$ is the set of first $j-1$ clicks.
Furthermore, notice that $p(\vec{c}_j \mid \mathcal{M}_i, \mathcal{C}_{1:j-1})$ is a measure of how relevant $\vec{c}_j$ is given model $\mathcal{M}_i$ in light of all past observations. 

This framework of competing models thus far is flexible to operate regardless of visualization design or dataset. Picking a likelihood function is the key part to fitting this framework into specific applications. For example, if our application involves a collection of text documents as the underlying dataset, we need to choose a likelihood function that describes user interactions with support over the set of text documents. If our application involves multidimensional data with continuous and discrete attributes, we need to choose a likelihood function that describes user interactions with support over all data points. In Section \ref{subsec_info_rel}, we narrow the scope of possibilities and consider a specific likelihood function that models information relevance in multidimensional datasets with continuous and discrete attributes.

\subsection{Model Definition for Multidimensional Data} \label{subsec_info_rel}

The focus of this section is on modeling how relevant a data point is to the user given their past interactions. This measure of relevance is denoted in Eq.\ \ref{eq_model_likelihood} as $p(\vec{c}_j \mid \mathcal{M}_i, \mathcal{C}_{1:j-1})$.
In more formal terms, we want to use the evolving interaction data to infer the user's \textbf{data objective,} which we define to be some user-determined probability distribution $p(\vec{x})$ over the data space with support over the items of interest. Points with higher values of $p(\vec{x})$ are considered more \textbf{relevant} to the session.
Our framework requires the models to be probabilistic (i.e. include information about uncertainty in outputs). 
There are multiple techniques for inferring the relevance of a data point to the user in light of past observations. Some examples include probabilistic classification techniques such as logistic regression and $k$-nearest neighbors model. This freedom in model choices makes our approach design-agnostic. Depending on the scenario and the dataset, practitioners may encode exploration patterns into appropriate models. In this work, we model the distribution of data points with which the user has interacted. This proof-of-concept approach provides us with a simple and interpretable method to explore the idea of competing models. Since the multi-dimensional datasets in this paper involve discrete and continuous dimensions, we utilize two kinds of parametric distributions: Gaussian and categorical. While the parametric nature of these distributions impose some restrictions (e.g. unimodality of the Gassian), their parameters make the distributions easier to interpret by simply observing the value of parameters. 

\paragraph{Continuous Dimensions}
The Gaussian distribution is commonly used for continuous variables. We utilize this family of distributions to model user's interest in each continuous dimension of the data.  For every model, $\mathcal{M}_i$, we construct a $|\mathbb{A}_{c, i}| + 1$ dimensional multivariate Gaussian to learn the distribution of observations for continuous attributes as well as the timestep (\textit{time}) of interactions, where $|\mathbb{A}_{c, i}|$ is the number of continuous dimensions involved in a given model $\mathcal{M}_i$ and an extra dimension is added to model the order at which observations arrive. Multivariate Gaussian distribution is parameterized by a mean vector $\Vec{\mu}$ and a covariance matrix $\Sigma$, representing center and spread of the distribution respectively. Since the value of these parameters $(\vec{\mu}, \Sigma)$ for a given set of observations $\mathcal{C}$ is unknown, we infer them thorough the Bayesian inference process which results in a closed form posterior predictive distribution function, $f_c(\vec{x} \mid \mathcal{C})$. This probability distribution function, $f_c$ is known as Student's t-distribution \cite{bishop:2006:PRML, murphy2007conjugate}. Moreover, the Bayesian inference for this distribution involves four hyper-parameters. We set these hyper-parameters so that the prior belief is \textit{uninformative} (i.e.\ contains no specific information about user exploration before interactions are observed). The derivation details of this posterior predictive distribution as well as the choice of hyper-parameters are further discussed in the supplementary material \footnote{\url{https://github.com/smonadjemi/competing_models}}. Finally, we normalize the probability density function $f_c$ across all data points in $\mathcal{D}$ to assign a probability value to the continuous dimensions of every point $\vec{x} \in \mathcal{D}$:
\begin{equation}
\label{pc-eq}
\begin{split}
    p_c(\vec{x} \mid \mathcal{M}_i, \mathcal{C}_{1:t}, time=t+1) = 
   \frac{f_c(\vec{x}[\mathbb{A}_{c, i}] \mid \mathcal{C}_{1:t}, time=t+1)}
   {\sum\limits_{\vec{x}' \in \mathcal{D}} {f_c(\vec{x'}[\mathbb{A}_{c, i}] \mid \mathcal{C}_{1:t}, time=t+1)}}
\end{split}
\end{equation}
In the equation above, $\vec{x}[\mathbb{A}_{c, i}]$ denotes the continuous dimensions of $\vec{x}$ that are relevant in model $\mathcal{M}_i$.  

\paragraph{Discrete Dimensions}
The categorical model is used to explain the probability of discrete events occurring. For an attribute domain with $K$ possible categories, the categorical model has a $K$-dimensional vector $\vec{\mu}$ which describes the probability of observing each of the $K$ choices. Since the value of $\vec{\mu}$ is unknown, we infer it via the Bayesian inference process. The Bayesian inference process for the categorical distribution relies on one hyper-parameter ($\vec{\alpha}$) representing the pseudocount for the prior. We set this hyper-parameter so that the prior is uninformative as demonstrated in Fig. \ref{fig:crimes-si-results} (i.e.\ contains no specific information about user exploration before interactions are observed). Let $d'$ be a categorical attribute with $K$ possible values. The posterior predictive of observing category $k$ given a set of interactions $\mathcal{C}$ is \cite{bishop:2006:PRML, tu2014dirichlet}:
\begin{align}
    \begin{split}
        f_{d'}(k \mid \mathcal{C}, \vec{\alpha}) &= \frac{\alpha_k + m_k}{\sum_{i=1}^{K}{(\alpha_i + m_i)}}
    \end{split}
\end{align}
where $\alpha_i$ is the pseudocount for category $i$ (from hyper-parameter $\vec{\alpha}$) and $m_i$ is the observation count for category $i$. We offer more details on this hyper-parameter and the derivation of closed-form posterior predictive in the supplemental material. We normalize the posterior predictive of each categorical distribution across all data points in $\mathcal{D}$ to assign a probability value to each discrete dimension of every point $\vec{x} \in \mathcal{D}$. For a discrete dimension $d'$ we have:
\begin{equation}
\label{pd-eq}
\begin{split}
    p_{d'}(\vec{x} \mid \mathcal{C}_{1:t}) = 
   \frac{f_{d'}(\vec{x}[d'] \mid \mathcal{C}_{1:t})}
   {\sum\limits_{\substack{\vec{x'} \in \mathcal{D} \\ \text{  s.t.  } \vec{x}[d'] = \vec{x'}[d']}}{1}}
\end{split}
\end{equation}
In the equation above, $\vec{x}[d']$ is the value of attribute $d'$ for data point $\vec{x}$. 

\paragraph{Combined Point Likelihood} Equations \ref{pc-eq} and \ref{pd-eq} provide us with probability distributions over continuous and discrete dimensions of the dataset respectively. The overall probability of any point $\vec{x} \in \mathcal{D}$ within a model $\mathcal{M}_i$ in light of past $t$ interactions $\mathcal{C}_{1:t}$ is:
\begin{equation}
\label{tbd-eq}
\begin{split}
    p(\vec{x} \mid \mathcal{M}_i, \mathcal{C}_{1:t}) = p_c(\vec{x} \mid \mathcal{M}_i, \mathcal{C}_{1:t}, \text{time}=t+1) \prod_{d \in \mathbb{A}_{d, i}}{p_d(\vec{x} \mid \mathcal{C}_{1:t})}
\end{split}
\end{equation}
%

This section provided a detailed explanation for initiating and maintaining a belief over a set of competing models. Specifically, section \ref{subsec:maintaining-a-belief-over-models} discussed the \textit{prior} and \textit{posterior} portions of equation \ref{eq_model_posterior}, and \mbox{section \ref{subsec_info_rel}} discussed the \textit{likelihood} portion of equation \ref{eq_model_posterior}. Going forward, we apply these concepts to detect exploration bias, predict future clicks, and summarize an analytic session.

\section{Applications of Competing Models} \label{sec:high_level_insight}
The ultimate goal of our technique is to make high-level and in-depth inferences by observing their low-level user interactions. More specifically, our technique takes a more comprehensive view of user modeling by enabling us to \textit{infer exploration bias}, \textit{predict future interactions}, and \textit{summarize an analysis session}. In this section, we use our  competing models framework (Section \ref{sec_competing_model_framework}) in order to gain insight into users.

\subsection{Exploration Bias Detection}

Bias often has a negative connotation; however, in reality bias can be desirable. In very simple cases, bias can indicate the criteria in a data space on which people make decisions \cite{wall2017warning}. Therefore, we categorize bias into two categories: intentional and unintentional. 
Intentional bias holds information about the users' search criteria and can accelerate the analysis process. Unintentional bias, on the other hand, arises from the user subconsciously avoiding some information and exploring others based on some personal factor. Unintentional bias leads to missing information. Regardless of the category, detecting biases in visualization systems can improve the quality of analysis and decision. 

Throughout a session, our framework maintains a posterior belief over the set of models \hbox{$\{\mathcal{M}_1, \mathcal{M}_2, ..., \mathcal{M}_{2^d}\}$} each of which represent biased exploration towards a subset of attributes. The normalized posterior belief, \hbox{$\{p(\mathcal{M}_1 \mid \mathcal{C}), p(\mathcal{M}_2 \mid \mathcal{C}), ..., p(\mathcal{M}_{2^d} \mid \mathcal{C})\}$}, informs us about which subset of attributes the user may be biased towards. Higher values of $p(\mathcal{M}_i \mid \mathcal{C})$ indicate higher chances of the user being biased towards attributes represented in $\mathcal{M}_i$. 

Using the normalized posterior belief over models, \hbox{$\{p(\mathcal{M}_1 \mid \mathcal{C}), p(\mathcal{M}_2 \mid \mathcal{C}), ..., p(\mathcal{M}_{2^d} \mid \mathcal{C})\}$}, we can use the law of total probability to calculate the chance of bias towards each individual attribute:
\begin{equation}
    \label{marginal-bias}
    p(\text{bias towards attribute }a \mid \mathcal{C}) = \sum_{\mathcal{M}_i \in \mathbb{M}_a}{p(\mathcal{M}_i \mid \mathcal{C})}
\end{equation}
where $\mathbb{M}_a$ denotes the subset of models in which bias towards attribute $a$ is assumed.

\subsection{Next Interaction Prediction}

Predicting which data points the user may interact with next opens opportunities to add features such as target assistance and target gravity to help users find the next most interesting data points in their exploration session \cite{ottley2019follow}. 
Anticipating next clicks given past clicks involves a time component. Adding time-steps as an additional dimension to our continuous model enables us to correlate clicks not only by their attributes, but also by the time-step at which they were clicked. For each of the data points, $x \in \mathcal{D}$, we calculate the probability of the user interacting with $x$ at the next time-step via:
\begin{equation}
\label{ncp-eq}
\begin{split}
    p(\vec{x} \mid \mathcal{C}_{1:t}, time=t+1) = \\
    \sum_{\mathcal{M}_i \in \mathbb{M}}{p(\vec{x} \mid \mathcal{M}_i, \mathcal{C}_{1:t}, time=t+1)p(\mathcal{M}_i \mid \mathcal{C}_{1:t})}
\end{split}
\end{equation}

The Bayesian approach for not losing any information by selecting one single model is \textit{Bayesian model averaging}, which we use in the equation above.

\subsection{Summarizing Analytic Session}
Using provenance data to summarize analytic sessions has been of interest in the visualization community \cite{xu2018chart, gratzl2016visual, heer2008graphical}. Due to our parametric choice of models in Section \ref{subsec_info_rel}, we show that those parameters can be used to represent a summary distribution from which the clicks were generated. 
For example, let $a$ be a continuous attribute. The updated parameters of the Gaussian distribution $(\mu_a, \sigma_a^2)$ after observing interactions can display a summary of what distribution of the values in attribute $a$ the user has explored. In the discrete case, the parameter of categorical distribution serves as a summary of observed interactions in a certain categorical domain. Section \ref{sec:validation} demonstrates this effect through figures.

\section{End-to-End Example}

In this section, we consider an end-to-end example with a small dataset to demonstrate our framework from Sections \ref{sec_competing_model_framework} and \ref{sec:high_level_insight}. Let $\mathcal{D}$ be a set of seven fictitious restaurants with \textit{location} and \textit{food type} being continuous and discrete attributes respectively. We encode each restaurant into a 3D vector of form $(latitude, longitude, type)$ to get:  

\begin{align*}
    \mathcal{D} = \{(0.35, 0.85, \text{Italian}), (0.8, 0.35, \text{Mexican}), (0.85, 0.1, \text{Persian}), \\ (0.7, 0.3, \text{Italian}), (0.15, 0.75, \text{Mexican}), (0.1, 0.05, \text{Persian}),\\ (0.9, 0.85, \text{Mexican})\}
\end{align*}

We assume that exploration based on latitude or longitude alone is likely unnatural, and combine the two to be a single 2D continuous attribute, \textit{location}. Hence, the set of continuous attributes is $\mathbb{A}_c = \{location\}$ and the set of discrete attributes is $\mathbb{A}_d = \{type\}$. Since the user may explore this dataset based on any combination of these attributes, we construct $2^{|\mathbb{A}_c| + |\mathbb{A}_d|} = 4$ models to represent exploration patterns (Table \ref{table:models_example}).
After two timesteps, we observe that the user has interacted with the following set of points (in the order listed):

\begin{align*}
    \mathcal{C} = \{(0.85, 0.1, \text{Persian}), (0.8, 0.35, \text{Mexican})\}
\end{align*}

The interaction type here is irrelevant, but for the sake of example we could assume it occurs through clicking, hovering, or adding to favorites.

\setlength{\tabcolsep}{6.3pt}
\begin{table}[!ht]
\centering
\caption{Example model space $\mathbb{M}$, where each model represents exploration based on a subset of attributes. \green{\faCheck} indicates an attribute is considered important to the exploration session and \red{\faTimes} indicates an attribute is not considered important to the exploration session.}
\begin{tabular*}{\linewidth}{lccccc}
\toprule
Model & \multicolumn{1}{c}{loc.} & \multicolumn{1}{c}{type} & $p(\mathcal{M}_i)$ & $p(\mathcal{M}_i \mid \mathcal{C}_{0:1})$ & $p(\mathcal{M}_i \mid \mathcal{C}_{0:2})$\\ \midrule
$\mathcal{M}_1$ & \multicolumn{1}{c}{\textcolor{sm-red}{\faTimes}} & \multicolumn{1}{c}{\textcolor{sm-red}{\faTimes}} & 0.25 & 0.08 & 0.03 \\ 
$\mathcal{M}_2$ & \multicolumn{1}{c}{\textcolor{sm-green}{\faCheck}} & \multicolumn{1}{c}{\textcolor{sm-red}{\faTimes}} & 0.25 & 0.12 & 0.10 \\ 
$\mathcal{M}_3$ & \multicolumn{1}{c}{\textcolor{sm-red}{\faTimes}} & \multicolumn{1}{c}{\textcolor{sm-green}{\faCheck}} & 0.25 & 0.29 & 0.16 \\ 
$\mathcal{M}_4$ & \multicolumn{1}{c}{\textcolor{sm-green}{\faCheck}} & \multicolumn{1}{c}{\textcolor{sm-green}{\faCheck}} & 0.25 & 0.51 & 0.70 \\ \bottomrule
\end{tabular*}
\label{table:models_example}
\end{table}

\begin{figure}[!ht]
    \centering
    \includegraphics[scale=0.19]{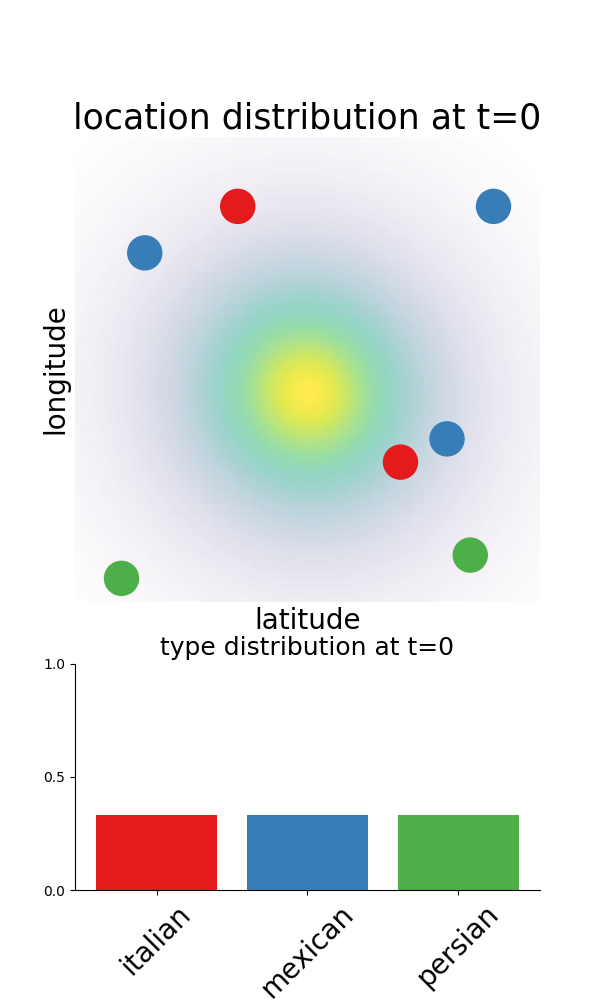}
    \includegraphics[scale=0.19]{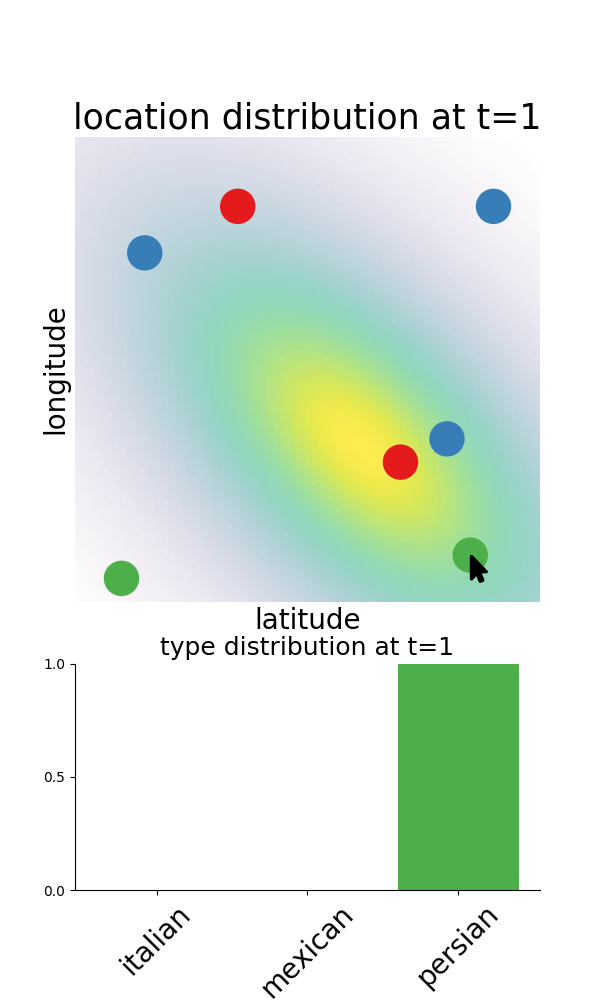}
    \includegraphics[scale=0.19]{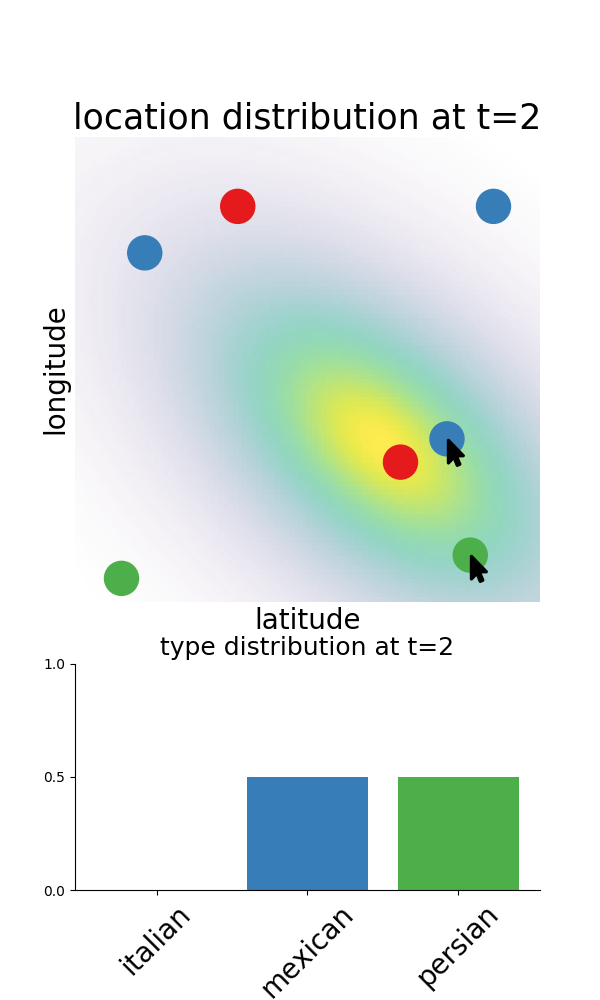}
    \caption{The inferred Gaussian and categorical distributions \textit{summarizing the first two interactions}. As clicks (denoted by \faMousePointer) arrive, our models get updated to represent user interactions.}
    \label{fig:ete_example}
\end{figure}

Before any interactions are observed, we are uniformly uncertain about which combination of attributes drive interactions ($p(\mathcal{M}_i)$ column in Table \ref{table:models_example}). As more clicks arrive, our posterior belief gains confidence that \textit{location} and \textit{type} attributes explain the interactions ($p(\mathcal{M}_4 \mid \mathcal{C}_{0:1}) = 0.51$ and $p(\mathcal{M}_4 \mid \mathcal{C}_{0:2})$ in Table \ref{table:models_example}). The posteriors (columns $p(\mathcal{M}_i \mid \mathcal{C}_{0:1})$ and $p(\mathcal{M}_i \mid \mathcal{C}_{0:2})$) are computed using a normalized probability mass function (Figure \ref{fig:ete_example}) as the likelihood function. 

In order to \textit{detect exploration bias} towards a certain attribute, we use Eq. \ref{marginal-bias} to compute the marginal probability of models involving that attribute. For example, the probability of bias towards location after observing two clicks is 0.80, since $p(\mathcal{M}_2 \mid \mathcal{C}_{1:2}) + p(\mathcal{M}_4 \mid \mathcal{C}_{1:2}) = 0.8$ and models $\{\mathcal{M}_2, \mathcal{M}_4\}$ both involve the \textit{location} attribute.

For \textit{next click prediction}, we compute the likelihood of every every point averaged over all models. For every point $x \in \mathcal{D}$, we compute:
\begin{align*}
    \sum_{\mathcal{M}_i \in \{\mathcal{M}_1, \mathcal{M}_2, \mathcal{M}_3, \mathcal{M}_4\}}p(x \mid \mathcal{M}_i, \mathcal{C}_{1:2}, time=3)p(\mathcal{M}_i \mid \mathcal{C}_{1:2})
\end{align*}
which results to $(\text{Mexican}, 0.15, 0.75)$ being the predicted next interaction (assuming no repeated clicks). More specifically, the order of points for being the next interaction is as follows:
\begin{align*}
    (\text{Mexican}, 0.8, 0.35) > (\text{Persian}, 0.85, 0.1) > (\text{Mexican}, 0.15, 0.75) > \\ (\text{Mexican}, 0.9, 0.85)  > (\text{Persian}, 0.1, 0.05) > (\text{Italian}, 0.7, 0.3) > \\ (\text{Italian}, 0.35, 0.85)
\end{align*}

\noindent Notice that \textit{location} and \textit{type} driving exploration (as inferred in Table \ref{table:models_example}) is reflected in the ordering above where the top choices for a third click (assuming no repeated clicks) are Mexican and Persian restaurants in order of their proximity to past interactions. Moreover, notice that the distributions in Figure \ref{fig:ete_example} start with uninformative priors when $t=0$, but update to \textit{summarize interactions} as they arrive ($t=1$ and $t=2$).

\section{Validation with User Study Datasets} \label{sec:validation}
We validate our proposed technique on three user study datasets. 
Each dataset seeks to highlight a unique aspect of our technique. First, we validate our technique using a user study dataset collected by Ottley et al. \cite{ottley2019follow} where users are asked to perform specific tasks using a map visualization of crimes in St. Louis. Since this user study dataset includes ground truths associated with each session, we can compare our outcomes of \textit{exploration bias detection} and \textit{next interaction prediction} with established baselines by Wall et al.\ \cite{wall2017warning} and Ottley et al.\ \cite{ottley2019follow} respectively.

Second, we demonstrate our technique's independence from visualization design by validating it on a user study dataset collected by Feng et al. \cite{feng2018patterns} where users freely interact with a visualization of S\&P 500 companies. While this dataset still involves a point-based visualization, it differs from the first dataset in that it is not a map-based visualization and the interaction technique is \textit{hovers} instead of \textit{clicks}. Considering that the ground-truth bias in these open-ended sessions are unknown, we are only able to compare the outcome of \textit{next interaction prediction} with the established baseline by Ottley et al.\ \cite{ottley2019follow}.

Lastly, we demonstrating our technique's independence from tasks by validating it on an open-ended user interaction session with the map of crimes in St. Louis, and compare our models' understanding of user exploration to the self-reported insight by the participant.

\subsection{Map of Crimes in St. Louis} \label{stl_crime_analysis}

\subsubsection{Ottley \textit{et al.} experimental setup}
Ottley et al.~\cite{ottley2019follow} describe a user study in which they capture mouse click data as participants interact with a map visualization of crimes in Saint Louis. In this experiment, participants were presented a set of 1,951 reported crimes from March 2017 visualized on an interactive map. Each instance was displayed as a dot with a position and a color indicating the \emph{location} and \emph{type} of the crime respectively. The map responded to user clicks by triggering a tooltip containing more details on the crime. The questions presented to the participants required them to search the map to find an answer based on the data:
\begin{enumerate}[nosep]
    \item Out of all the cases of Homicide, one case differs from the other cases with regard to time. What is the time of that case?
    \item How many cases of arson occur during PM?
    \item There are four types Theft-Related crime in the red shaded region: Larceny, Burglary, Robbery and Motor Vehicle Theft. Count the number of cases of Robbery in the red shaded region.
    \item There are two types of Assault: Aggravated and Non-Aggravated assault. Count the number of Non-Aggravated Assault in the red shaded region
    \item Count the number of crimes that occur during 7:00 AM - 12:30 PM in the red shaded region.
    \item Count the number of crimes during AM in the red shaded region. 
\end{enumerate}

\begin{figure}[!t]
    \centering
    \includegraphics[scale=0.26]{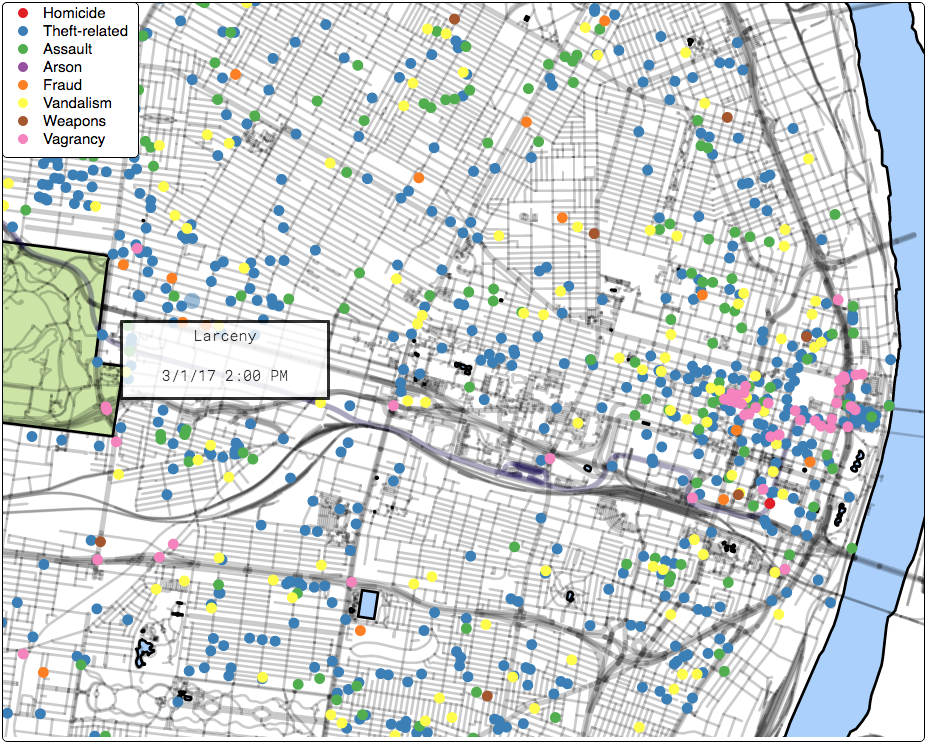}
    \caption{The map of St. Louis crimes used by Ottley et al. \cite{ottley2019follow}. The position of the dots indicate the \textit{location} of the crime, and the color of the dots indicate the \textit{type} of crime. Users explored the data by \underline{clicking} on the dots and observing more information in a tooltip.}
    \label{fig:stl_crime_map}
\end{figure}

\begin{table}
\centering
\caption{Number of participants and ground truth exploration biases for each category of tasks in the Ottley et al. experiment \cite{ottley2019follow}}
\begin{tabular*}{\linewidth}{lp{3.5cm}l}
\toprule
Task Category & Ground-Truth Bias & \# of Participants \\ \midrule
Geo-Based              & \textit{latitude, longitude}      & 28            \\ 
Type-Based             & \textit{type}                      & 23           \\ 
Mixed                  & \textit{latitude, longitude, type}       & 27     \\ \bottomrule
\end{tabular*}
\label{table:crime_biases}
\vspace{-5mm}
\end{table}

The six questions above were categorized into three groups of tasks: \textit{geo-based} (search specific location, questions 5-6), \textit{type-based} (search for specific type of crime, questions 1-2), and \textit{mixed} (search for a particular type of crime in a specific region of the map, questions 3-4). Table \ref{table:crime_biases} summarizes the number of participants for each group of tasks.

\begin{figure*}
    \centering
    \includegraphics[scale=0.34]{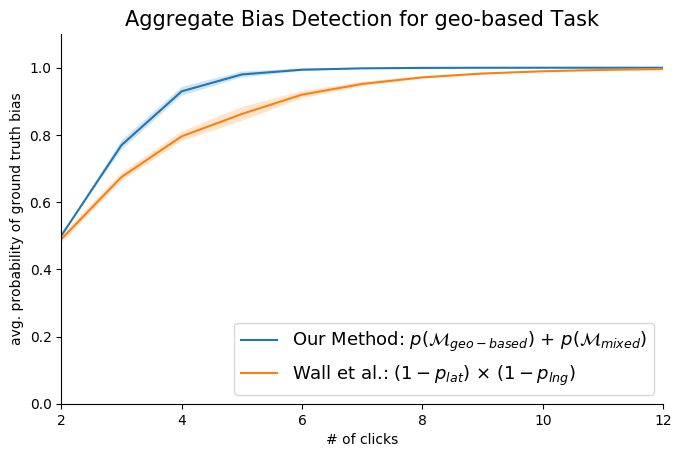}
    \includegraphics[scale=0.34]{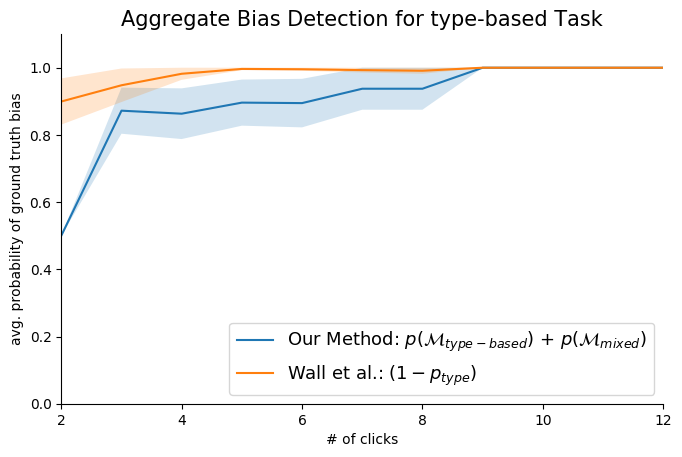}
    \includegraphics[scale=0.34]{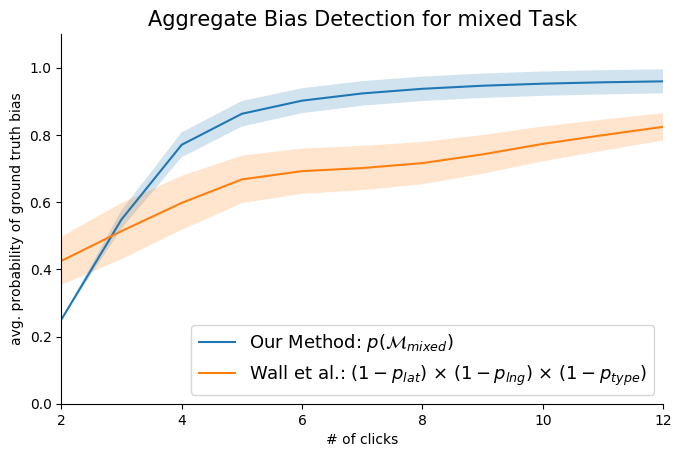}
    \caption{Results of \textbf{\textit{exploration bias detection}} on user study dataset from the Ottley et al. experiment (Section \ref{stl_crime_analysis}). The participants were asked to interact with a map visualization of crimes in St. Louis in order to answer geo-based questions  \textit{(based on latitude and longitude)}, type-based questions \textit{(based on type)}, and mixed questions \textit{(based on latitude, longitude, and type)}. After 12 clicks, our method significantly outperformed Wall et al. on the geo-based and mixed tasks according to a paired t-test with p-value $1.5 \times 10^{-7}$ and $0.003$, respectively. Wall et al. outperformed our method on the type tasks with p-value 0.1. The left graph shows that our method detects exploration bias towards \textit{location} quicker and more confidently than the baseline. The middle graph shows that our technique detects exploration bias towards the \textit{type}, however, it takes longer than the baseline to gain confidence in the level of bias (discussed in Section \ref{res:stl-crime-analysis}). The right graph shows that our method detects bias towards \textit{location and type} quicker and more confidently than the baseline. The shaded region in all three graphs represent the standard error among 28 geo-based sessions, 23 type-based sessions, and 27 mixed sessions.}
    \label{fig:crimes-bias-results}
\end{figure*}

\begin{figure*}
    \centering
    \includegraphics[scale=0.34]{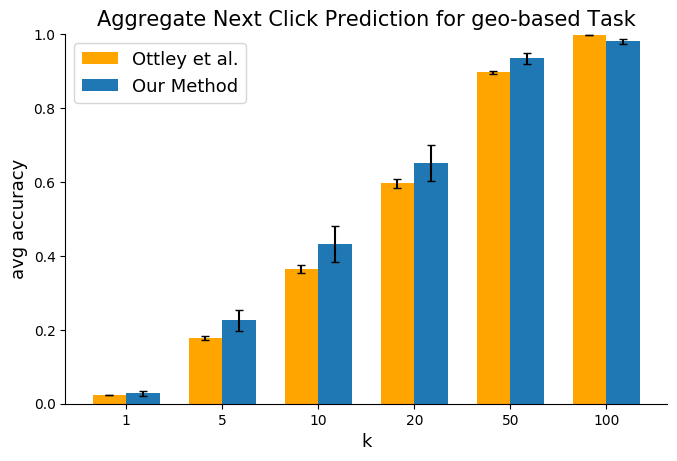}
    \includegraphics[scale=0.34]{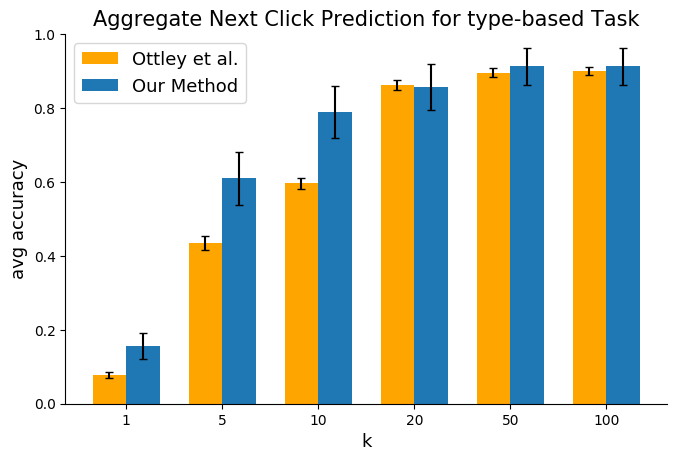}
    \includegraphics[scale=0.34]{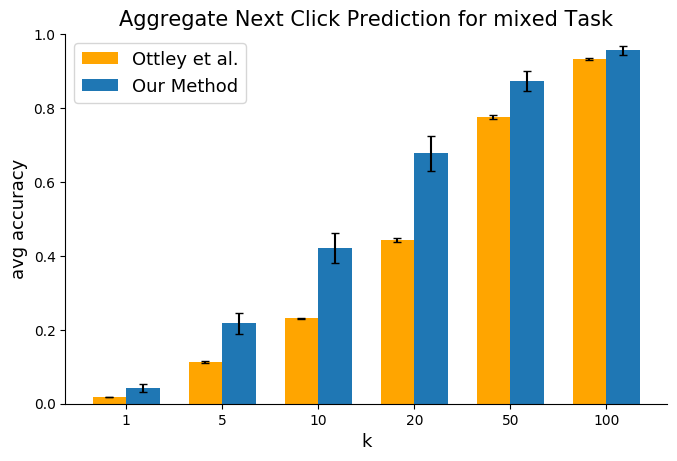}
    \caption{Results of \textbf{\textit{next click prediction}} on user study dataset from the Ottley et al. experiment (Section \ref{stl_crime_analysis}). Our method performs within the margin of error of the previously studied hidden Markov model in all categories, while outperforming it in most cases. The error bars in all three graphs represent the standard error among 28 geo-based sessions, 23 type-based sessions, and 27 mixed sessions. }
    \label{fig:crimes-ncp-results}
\end{figure*}

\begin{figure*}
    \centering
    \includegraphics[scale=0.31]{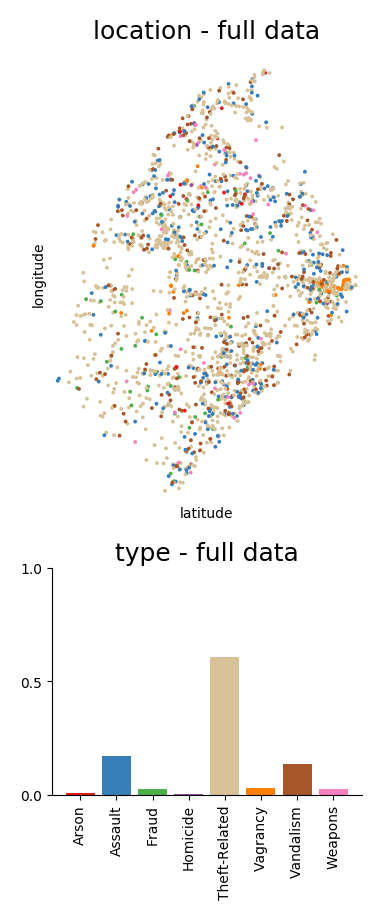}
    \unskip\ \vrule\ 
    \includegraphics[scale=0.31]{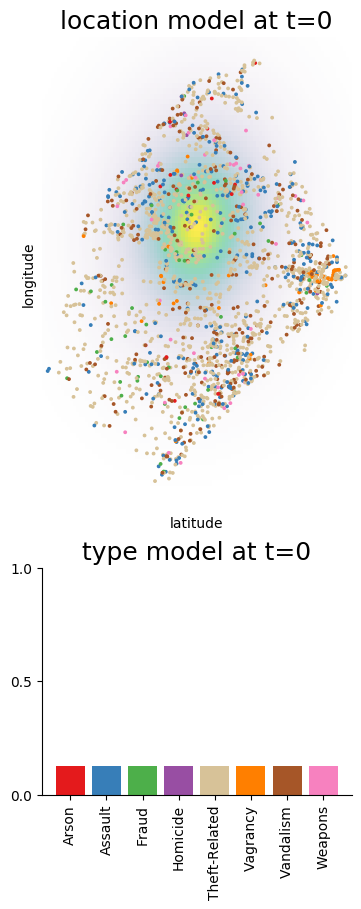}
    \includegraphics[scale=0.31]{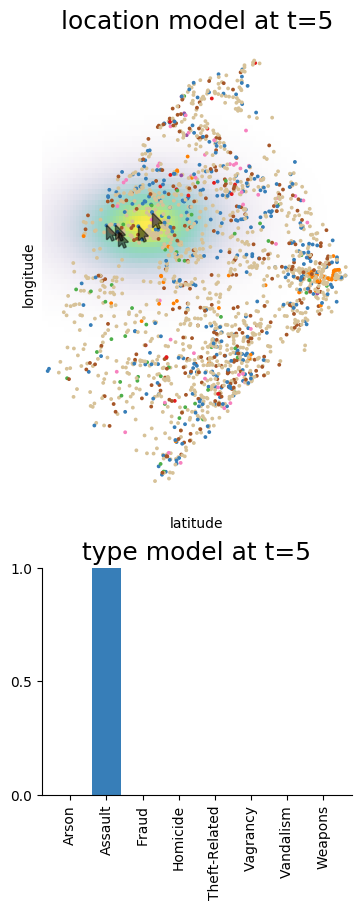}
    \includegraphics[scale=0.31]{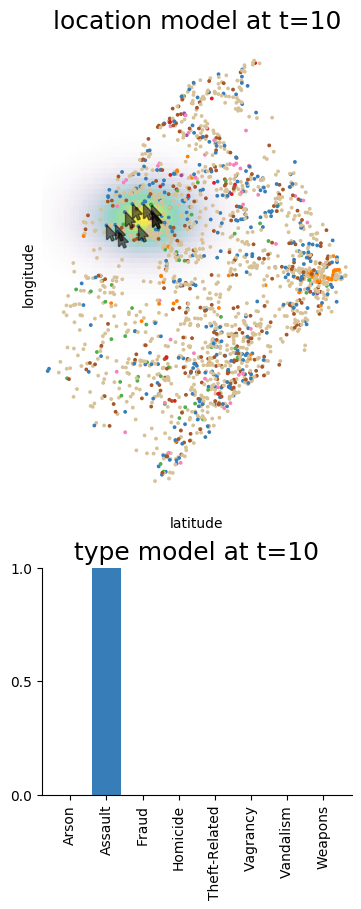}
    \includegraphics[scale=0.31]{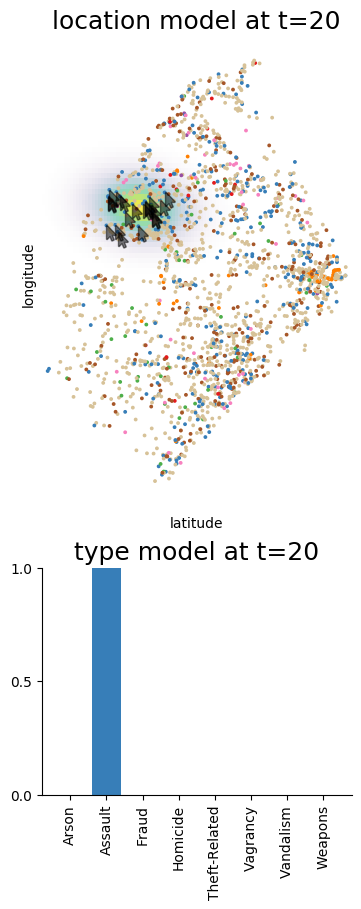}
    \includegraphics[scale=0.31]{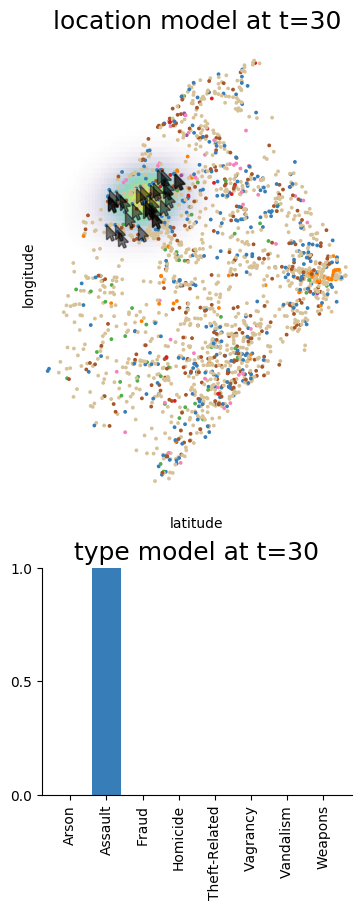}
    \caption{An instance of \textbf{\textit{summarizing analytic sessions}}, where the user is performing mixed interactions \textit{(based on location and type)} in order to answer question 4 from Section \ref{stl_crime_analysis}. The left-most graph shows the distribution of the full crime dataset presented to participants. At $t=0$ before any clicks have been observed, our models reflect our prior. As clicks (denoted by \faMousePointer) arrive, our models get updated to represent user interactions. Notice the significant difference in \textit{type model} from $t=0$ to $t=5$. This is because our pseudo-count for the prior is small, making the prior insignificant in comparison to observed user clicks.}
    \label{fig:crimes-si-results}
\end{figure*}

\subsubsection{Results}\label{res:stl-crime-analysis}
In our first set of analysis, we consider \textit{\textbf{exploration bias detection}} based on the ground-truths presented in Table \ref{table:crime_biases}, and we compare the output of our technique from Eq. \ref{marginal-bias} with the equivalent baseline metric. As discussed in Section \ref{background}, Wall et al. \cite{wall2017warning} propose an \textit{attribute distribution} metric to measure how biased an exploration session is towards any particular attribute. They define this metric to be the complement of observing the set of interactions assuming the distribution of attributes within the full dataset and interaction dataset are the same. They specifically suggest the non-parametric KS test for continuous attributes and Chi-Square test for discrete attributes. For each attribute, these tests take two sets of 1-dimensional sets as inputs (full data and interaction data), and decide if the two sets are from the same distribution. Let $p_{a}$ be the output of a KS or Chi-Square test for attribute $a$; then, the \textit{attribute distribution} metric is defined as $b_{Ad}(a) = 1 - p_{a}$. Note that higher values of $b_{Ad}(a)$ correspond to more exploration bias towards attribute $a$. For tasks with bias towards more than one attribute, we compute the product of this \textit{attribute distribution} metric to represent the conjunction of biases. A comparison between our algorithm's bias detection and the baseline is shown in Figure \ref{fig:crimes-bias-results}. 
After 12 clicks, our method significantly outperformed Wall et al. on the geo-based and mixed tasks according to a paired t-test with p-value $1.5 \times 10^{-7}$ and $0.003$, respectively. Wall et al. outperformed our method on the type tasks with p-value 0.1.
Upon further investigation, we concluded that the better performance of baseline in type-based sessions is explained by the characteristic of corresponding type-based questions. Questions 1 and 2 ask participants to explore under-represented categories of the data (i.e. Homicide and Arson), and Chi-Square tests can detect that bias quicker. On the other hand, our technique outperforms the Chi-Square baseline when detecting bias towards over-represented categories. While the user study datasets did not include sessions to reflect this effect, we verified them using synthetic sessions from the crime dataset.

In our second set of analysis for this dataset, we consider \textit{\textbf{next click prediction}} and compare the outcome of our technique with the baseline from Ottley et al. \cite{ottley2019follow}. This baseline is based on the idea that users click on items that are within a close proximity from each other. The implementation of this baseline involves iteratively sampling particles from a current belief, using the observation model to re-weight the particles based on observed clicks, and ranking the datapoints according to the weight and proximity of particles at each timestep. Then, the set of top-$k$ candidates are selected to represent a prediction set for the next datapoint with which the user may interact. In order to be consistent with the existing work, we follow a similar approach in our technique: after averaging our competing models to compute the predictive posterior of each data point being the next observation as outlined in Eq. \ref{ncp-eq}, we sort them based on ranking and pick the top-$k$ data points as the set of candidates for next interaction. At each time-step, we determine if our prediction set included the next click and compute the rate of success in this prediction process. A comparison between our algorithm's next click prediction and the baseline is shown in Figure \ref{fig:crimes-ncp-results}. Across all three tasks, our technique outperforms this baseline for the majority of $k$ values and performs within the margin of error for the rest. Anticipating future data points with which the user may interact with has implications in designing more intelligent visualization systems which understand user goals and assist them in finding the information they desire. We discuss more of this topic in Section \ref{future-work}.

The final goal of our technique is to \textit{\textbf{summarize analytic sessions}}. Since we chose parametric models to represent user interactions, summarizing a session is just a matter of observing the value of parameters at each time-step. Figure \ref{fig:crimes-si-results} shows the progression of our algorithm from as new interactions arrive. By only looking at the summary of interactions through time, we can easily map the session to question 4, where the user is exploring Assault cases in a particular region of \hbox{St. Louis}.

\begin{figure}[!t]
    \centering
    \includegraphics[scale=0.078]{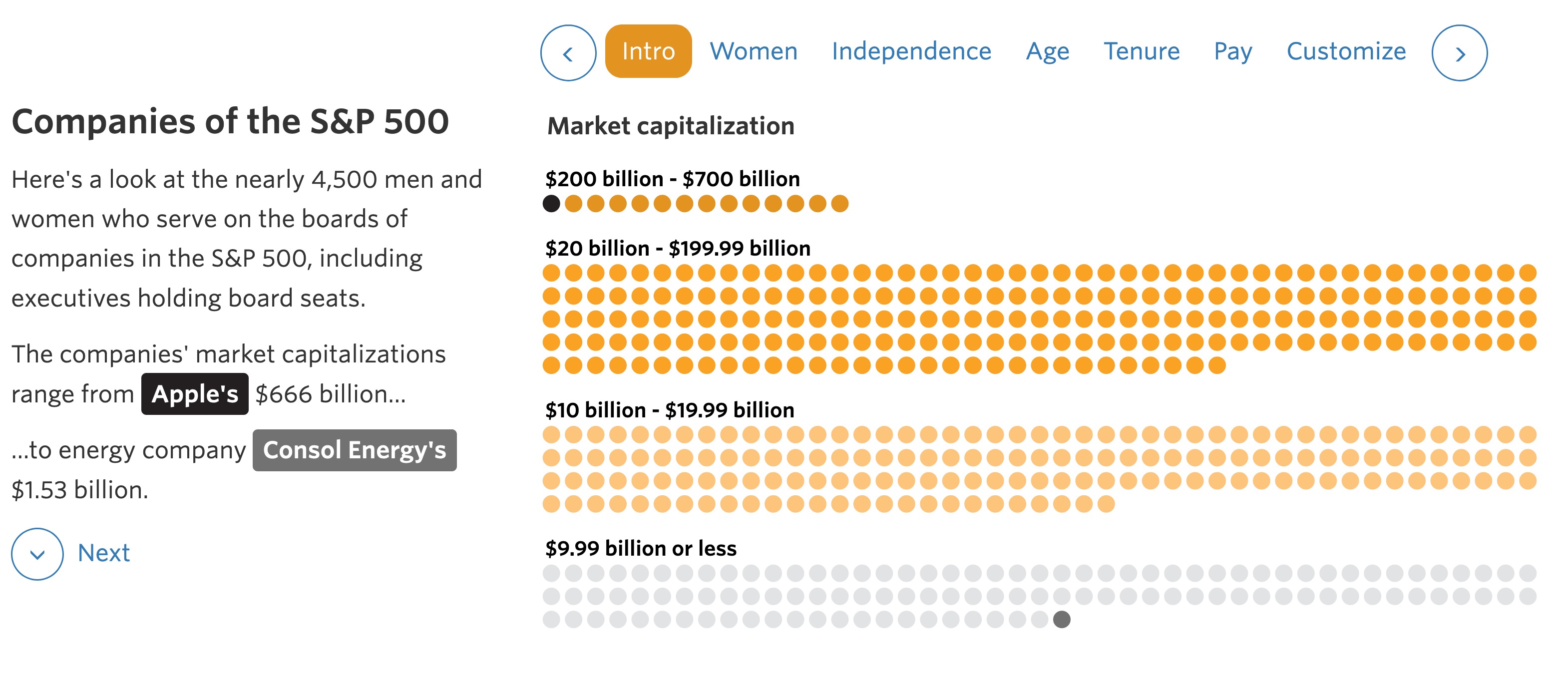}
    \vspace{-5mm}
    \caption{``Inside America's Boardrooms'' a multi-section visualization published in the Wall Street Journal presenting data on leaders of S\&P 500 companies \cite{wsjboardrooms}. Users from the Feng et al. experiment \cite{feng2018patterns} interacted with this visualization by \textit{hovering} over different companies and observing more information about the leaders of the companies in a tooltip.}
    \label{fig:boardrooms_wsj_view}
    \vspace{-5mm}
\end{figure}

\subsection{S\&P 500 Boardrooms} \label{boardrooms-analysis}
Thus far in our validation, we have relied on a map visualization with which users have to interact through clicks in order to accomplish pre-defined tasks. In this part of our validation, we consider a multi-section visualization of S\&P 500 board of directors published by Wall Street Journal \cite{wsjboardrooms}. The underlying data for this visualization has one discrete and six continuous attributes. A tooltip containing these attributes appears on hover. A view of this visualization is shown in Figure \ref{fig:boardrooms_wsj_view}. 

\subsubsection{Feng \textit{et al.} experimental setup}

Feng et al. \cite{feng2018patterns} collected user interaction sessions, where users freely interact with point-based elements in a visualization and search for keywords. Specifically, they use the visualization called "Inside America's Boardroom" published in the Wall Street Journal (Fig. \ref{fig:boardrooms_wsj_view}). This underlying data for this visualization contains 7 dimensions: \textit{market capitalization, ratio of unrelated board members, ratio of female board members, average age of board members, average tenure, median pay, and industry group}.
As opposed to the Ottley et al. experiment, participants in this study were asked to perform an open-ended exploration on the dataset. The recorded sessions contain a sequence hovers on different data points and the duration of each hover. While this user study was designed to study the impact of having search capabilities in visualizations, we utilize the hover data to validate our technique for predicting future hovers. Since hovers can inherently be noisier than clicks due to unintentional hovers while going from one data point to another, we filtered the sessions to only include hovers that lasted for over one second. This will eliminate most unintentional hovers that occurred in transition of going from one data point to another. Moreover, we filtered the data further to include only those with more than 3 hovers. Before filtering, there were 41 sessions with avg.\ 391.19 hovers per sessions (SD 237.75). After filtering, we had 39 sessions with avg.\ 26.83 hovers (SD 18.95) and at least 3 hovers per session.

\subsubsection{Results}

\begin{figure}[!ht]
    \centering
    \includegraphics[scale=0.45]{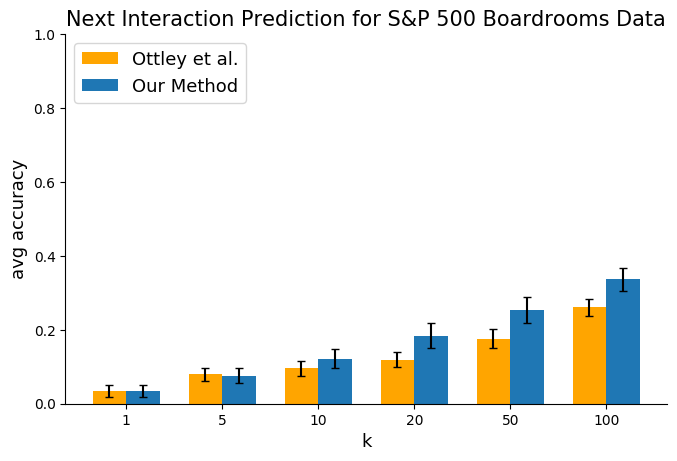}
    \vspace{-5mm}
    \caption{Average accuracy of predicting the next hover in the set of top-$k$ candidates for boardrooms dataset. There are 39 sessions, and prediction starts after the third click for consistency with the baseline from Ottley et al. \cite{ottley2019follow}. The error bars indicate the standard error of accuracy across all sessions. There are 500 data points in the visualization, making $k=50$ to be 10\% of the dataset. Our technique outperforms or performs within margin of error in comparison to the baseline.}
    \label{fig:boardrooms-ncp-res}
\end{figure}

In our analysis of this user study, we consider \textbf{\textit{next hover prediction}}. We modify the Ottley et al. baseline from Section \ref{stl_crime_analysis} to include all seven attributes of this dataset. Our technique creates a space of $2^7$ models (corresponding to all possible subsets of the 7 dimensions), where each model represents biased exploration towards a certain subset of attributes. Furthermore, we make an adjustment to the prior belief over the model space in order to penalize models that encompass bias towards a large subset of attributes. In Section \ref{subsec:maintaining-a-belief-over-models}, we suggested the uniform prior belief $p(\mathcal{M}_i)=1/2^d$, however, here we use $p(\mathcal{M}_i) \propto 1/(d+1)$, where $d'_i$ is the number of attributes represented in model $\mathcal{M}_i$. Figure \ref{fig:boardrooms-ncp-res} shows that our technique outperforms the established baseline for higher values of $k$ and performs within the margin of error for lower $k$.

\subsection{Open-ended tasks with Map of Crimes in St. Louis} \label{subsec:open-ended-validation-stl}

In our final phase of validation, we reconsider the map of crimes in \hbox{St. Louis}. The experiments in Section \ref{stl_crime_analysis} involved narrowly defined tasks. In this case-study, we have selected one analytic session where the user freely interacted with the visualization and self-reported an insight they had. The main purpose of this final case study is to highlight how the self-reported insight relates to our model's summary of session. 

\begin{figure}[!ht]
    \centering
    \includegraphics[width=\linewidth]{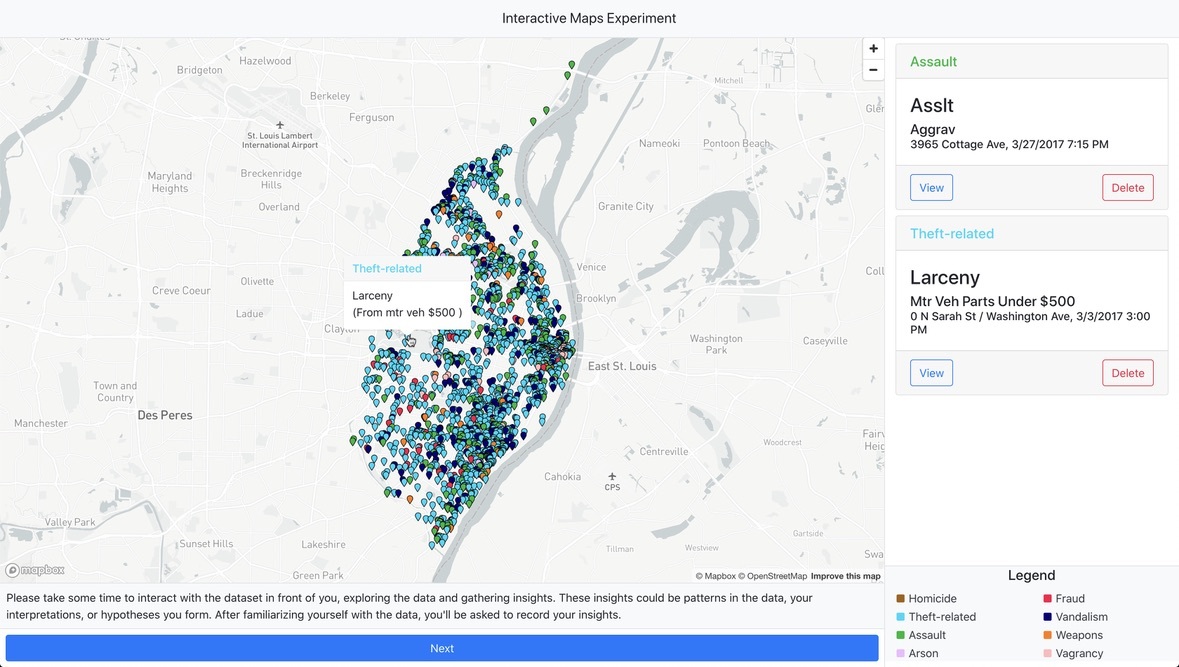}
    \vspace{-6mm}
    \caption{The experiment interface from Kern et al. \cite{kern2020effects}. Crime cases were visualized on an interactive map, where the colors indicate the type of crimes. Users were able to zoom, pan, hover, and click. While hovering on dots, a tooltip with more details opens. When a dot is clicked, the crime case is added to the sidebar as shown above. }
    \label{fig:stl_kern_view}
    \vspace{-5mm}
\end{figure}

\subsubsection{Kern \textit{et al.} experimental setup}

In this crowdsourced experiment, Kern et al. \cite{kern2020effects} propose an alternative experimental design in which participants freely interact with a map visualization of crimes in \hbox{St. Louis} and report their insights. Figure~\ref{fig:stl_kern_view} shows the visual interface for this experiment, hovers trigger a tooltip with more details and clicks add the data point to the side bar. In this paper, we used clicks from one session that had $\geq 10$ click events. Before the participants start the experiment, they are asked to familiarize themselves with the interface and functionalities. We use recorded clicks from this experiment to study how our model's understanding towards the user compares with the self-reported insight.

\subsubsection{Results}

\begin{figure}[!t]
    \centering
    \includegraphics[scale=0.3]{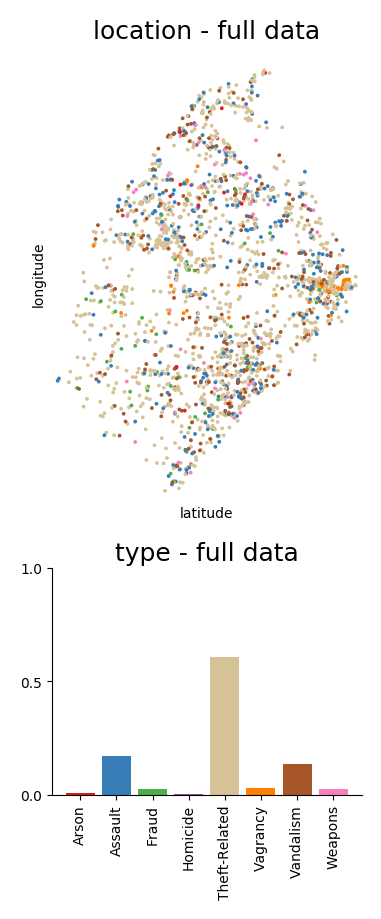}
    \unskip\ \vrule\ 
    \includegraphics[scale=0.3]{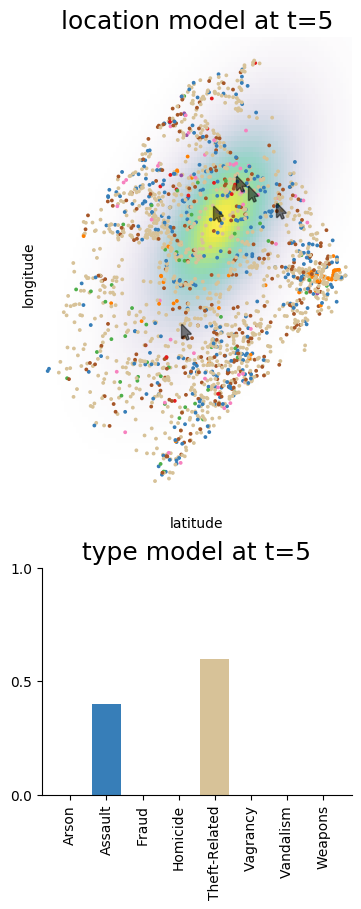}
    \includegraphics[scale=0.3]{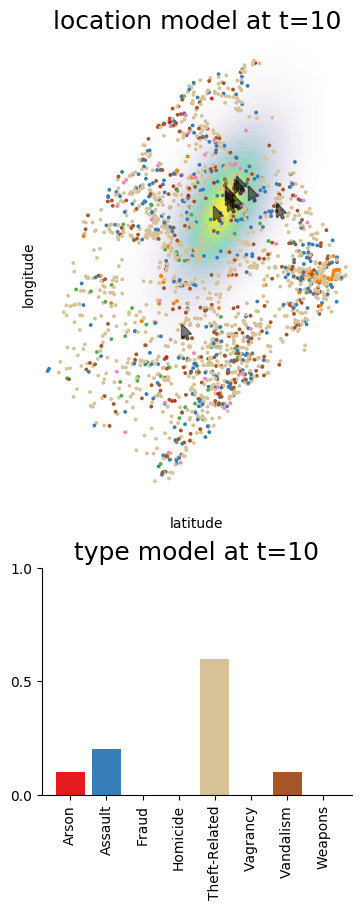}
    \caption{The summary of a session in which the user is freely interacting with a map visualization to report their insights. The left-most graph shows the distribution of the full crime dataset presented to the participant. As clicks (denoted by \faMousePointer) arrive, our models get updated to represent user interactions. Notice that type model at time $t=10$ does not significantly differ from the type distribution in the full data set, resulting in no detection of exploration bias towards \textit{type}.}
    \label{fig:crimes-openended-si-results}
    \vspace{-5mm}
\end{figure}

After observing six clicks in this session, our model posterior inferred the user is exploring based on only location (geo-based) with probability $>0.8$. By the time eight clicks were observed, our model was more certain on this task being geo-based with probability $>0.95$. Notice that this finding is trivially justified in Figure \ref{fig:crimes-openended-si-results}, where the distribution of \textit{type} attribute in the interaction set is very similar to the full data distribution (hence no bias towards type), but the distribution of $location$ attribute in the interaction set is narrow compared to the full dataset. 
Our model's understanding of the user reflects their self-reported insight: 
\textit{``Two cars were stolen in the same week from the 3900 block of Miami St.''}
We hypothesize that the absence of bias towards \textit{type} is due to the user looking for an insight among different \textit{type}s of crimes in the session.

\section{Discussion} \label{discussion}

The novelty of our technique is that we encode different exploration patterns as models and use the Bayesian model selection to maintain a belief over all possible exploration patterns. To demonstrate how a set of competing models can uncover user exploration patterns, we designed our model space so that each model represents information relevance based on a subset of dimensions in data. 

One of the immediate insights from our results is that posing user exploration modeling as a Bayesian model selection problem results in better performance in \textit{exploration bias detection} and \textit{next interaction prediction} in most cases. In particular, our choice of likelihood function from Section \ref{subsec_info_rel} resulted in improved performance for sessions in which the users were asked to explore a specific subset of the data (Figures \ref{fig:crimes-bias-results}, \ref{fig:crimes-ncp-results}, and \ref{fig:crimes-si-results}). In the case of bias detection for type-based tasks (Fig. \ref{fig:crimes-bias-results}, middle), the baseline outperformed our technique. After further investigation, we concluded that this behavior was due to the Wall et al. \cite{wall2017warning} baseline taking into account the overall data distribution whereas our likelihood function does not take into account the overall data distribution. An alternative choice of likelihood function that takes the underlying data distributions into account may result in improvements.

For open-ended tasks, we believe there could be more expressive choices of models to learn multi-modal distributions or more complex decision boundaries. Although our technique outperformed the existing baseline for open-ended tasks (Figure \ref{fig:boardrooms-ncp-res}), we believe a more expressive choice of models and a more specific user study to prevent users from unintentional hovers would be beneficial.
Another limitation of our work is that interactions that manipulate the underlying dataset are not supported as we assume our dataset $\mathcal{D}$ is constant throughout the session. In other words, we could consider interactions such as drag/drop to learn which subset of the data one may interact with, however, further investigation is needed to build models that learn about the context of re-positioning and make more complex predictions (such as where should a given point move given past movements).

The technique presented in this paper takes the visualization community a step closer to modeling users in scenarios where there are multiple possible exploration strategies, analysis goals, or personal preferences. By representing each possibility as a model and updating the belief as evidence is observed (through interactions), we can enable intelligent machine response in visualization systems. However, further work is required to address some of the shortcomings of our study. We have identified three main areas for further exploration, which we discuss in the next section.

\section{Future Directions} \label{future-work}
Framing user exploration modeling as a Bayesian model selection problem opens a new set of research opportunities. In particular, we can utilize recent advancements in Bayesian machine learning (i.e. active learning, active search, etc.) to further improve the collaboration between human and computer in visual analytics systems. In this section, we highlight four specific areas for future investigations.

\paragraph{Different Model Spaces}
Visual analytics research is constantly evolving to study new human-related factors that impact how we use interactive visualization systems. This evolution in user modeling alongside our competing models framework provide an opportunity to infer which human-related factor is impacting an individual session the most and move towards more individualized visual analytics systems.

\paragraph{Sampling Methods for More Efficiency}
Throughout this work, we compute exact values of model posteriors. While this was not prohibitive for our largest model space ($2^7$ models in Section \ref{boardrooms-analysis}, where there were 7 attributes), we recognize computing exact posteriors is not always feasible. This could be due to the unavailability of posterior predictive distributions in closed form or the need for low latency in real-time systems. In such scenarios, sampling methods may be used to approximate the posterior distribution. 

\paragraph{Active Learning to Mitigate Information Bubble}
Active learning is the idea behind an algorithm that queries an oracle in order to learn. In the scope of our work, this concept can be utilized to confirm detected biases with the user in order to avoid unintentional biases (through \textit{exploration}) and mitigate information overload by focusing on relevant information (through \textit{exploitation}). The mechanics of effectively querying users and updating the visualization view are also left for future studies. This line of research can lead to building \textit{automatic filtering} features to mitigate information bubbles created by unintentional bias.

\section{Conclusion} \label{conclusion}

We began this paper by claiming that interactive visualizations alone cannot fully support humans in analyzing large datasets and making informed decisions. There are numerous human factors such as limited cognitive capacity and personal biases that have roles in hindering effective analysis of data.
As the visual analytics research community investigates more about human factors related to data analysis, we need to build intelligent visual analytic systems who understand users through the lens of their low-level interactions and ultimately improve the visual analytic experience. 
In order to enable machine teammates to make high-level inferences by observing low-level interactions, we proposed \textit{competing models}: a technique which enumerates a set of human-related possibilities as models and then utilizes iterative Bayesian updating to learn about users as interactions are observed.
To narrow our focus, we outlined the process of creating and maintaining a set of competing models, and described how visual analytics researchers can use this idea to gain a deeper insight into user exploration. 
In the process of validating our technique, we demonstrated that by building models on the underlying dataset in visualizations and then updating said models as interactions are observed, we can uncover exploration biases and predict future interaction. In particular, we reached high rates of accuracy for \textit{next interaction prediction} when the participants were asked to perform specific tasks. In open ended scenarios, we saw a lower overall performance while still outperforming the baseline. This observation calls for more investigation on modeling open-ended sessions, which is inherently a more difficult objective.

\acknowledgments{
The authors wish to thank Sunwoo Ha for assisting in data preparation. Lane Harrison, Mi Feng, and Adam Kern for sharing their data. Emily Wall for her conversation on the bias detection metric. This material is based upon work supported by the National Science Foundation under grant numbers 1755734, 1845434, and 1940224.}

\bibliographystyle{abbrv}

\bibliography{paper}
\end{document}